\documentstyle[12pt]{article}
\normalsize
\def\sp{~~~~~}

\def\a{\alpha}
\def\b{\beta}

\def\d{\delta}
\def\e{\epsilon}

\def\g{\gamma}

\def\m{\mu}
\def\n{\nu}
\def\o{\omega}
\def\p{\pi}

\def\r{\rho}
\def\s{\sigma}

\def\G{\Gamma}

\def\cf{{\cal F}}

\def\cl{{\cal L}}

\def\rt{\rightarrow}

\def\bar#1{\overline{#1}}

\def\Hat#1{\rlap{\kern.10em$\widehat{\phantom G}$}#1}
\def\HAt#1{\rlap{\kern.05em$\widehat{\phantom G}$}#1}

\def\cap#1{\rlap{\kern.1em$\widehat{\phantom{G\vrule height.8em}}$}#1{}}
\def\Cap#1{\rlap{\kern.05em$\widehat{\phantom{G\vrule height.8em}}$}#1{}}

\let\oldtheequation=\theequation
\def\doteqs#1{\setcounter{equation}{0}
            \def\theequation{{#1}.\oldtheequation}}
\newcounter{sxn}
\def\sx#1{\addtocounter{sxn}{1} \bigskip\medskip \goodbreak
\noindent{\large\bf\centerline{\thesxn.~~#1}} \nobreak \medskip}
\def\sxn#1{\sx{#1} \doteqs{\thesxn}}

\newcounter{axn}

\def\br{\begin{thebibliography}{abc}}
\def\er{\end{thebibliography}}

\def\be{\begin{equation}}
\def\ee{\end{equation}}
\date{}

\begin{document}
\bibliographystyle{unsrt}
\footskip 1.0cm
\thispagestyle{empty}
\setcounter{page}{0}
\begin{flushright}
SU-4228-545\\
hep-ph/9309301 \\
August 1993\\
\end{flushright}
\vspace{10mm}
\centerline{\Large \bf RESONANT AND NON-RESONANT PIECES}
\vspace{5mm}
\centerline{\Large \bf OF THE $D\rt \bar{K}\pi$}
\vspace{5mm}
\centerline {\Large \bf SEMILEPTONIC TRANSITION AMPLITUDE}
\vspace*{8mm}
\centerline {\large J. Schechter, A. Subbaraman,\footnote{Address after
September 1, 1993, Physics Department, University of California, Irvine CA
92717} \small and \large S. Surya}
\vspace*{5mm}
\centerline {\it Department of Physics, Syracuse University,}
\centerline {\it Syracuse, NY 13244-1130}
\vspace*{25mm}
\normalsize
\centerline {\large Abstract}
\vspace*{5mm}

We compare the resonant and non-resonant contributions in various regions of
phase space for the $D\rt \bar{K}\p$ semileptonic transition amplitude,
computed in a chiral model which incorporates the heavy quark symmetry.
Remarks on the significance for experiment and for chiral perturbation theory
are made.

\newpage

\baselineskip=24pt
\setcounter{page}{1}

\sxn{INTRODUCTION}

In this note we shall examine the weak hadronic current matrix element for the
decay $D^0\rt K^-\p^0 e^+\n_e$ using a chiral Lagrangian which incorporates the
heavy quark
symmetry.  Previous works have treated this process in a chiral model which
includes only light pseudoscalars [1] and in chiral models with both light
pseudoscalars and light vectors present but with the approximation that the
decay be replaced by $D^0 \rt K^{*-}e^+\n_e$  [2-6].  Here we will consider
{\it both\/} contributions
together.  This is interesting, of course, in its own right.  It also holds
some interest for the question of what is the best way to incorporate vector
mesons
in the chiral perturbation theory program [7].  It may actually be easier to
investigate this question in the framework of chiral light-heavy interactions
rather than light-light interactions since the ``heavy end'' might eventually
be under better control.  For  the present process we find that there is no
region of phase space in which the light vector $K^*$ piece does not make a
non-negligible contribution and that it is typically very dominant.  This
perhaps suggests the adoption of a framework in which light pseudoscalars and
light vectors are
treated together from the beginning.

The detailed points include the discussion of the way the chiral theorem is
maintained in the appropriate unphysical limit and the treatment of the phase
space kinematics for the hadronic matrix element.  The transition amplitude is
given in
Section 2 (see also the Appendix) after our notation has been introduced.
Section 3 contains the kinematics and the comparison of the resonant and
non-resonant pieces of the amplitude in various regions of phase space.  Some
caveats and remarks on the experimental aspect of $K^*$ dominance are given in
Section 4.
\newpage

\sxn{Decay Amplitudes}

We are interested in the chiral invariant interactions of both the light
pseudoscalar nonet $\phi(x)$ and the light vector nonet $\r_\m(x)$ with the
heavy
meson field.  We shall follow the notation of Ref. [3];  other treatments
include Refs. [2, 4, 5, 6].  The chiral interactions involving only the light
pseudoscalars were discussed in Ref. [8].  Using the ``heavy field'' $H(x)$
which contains both heavy pseudoscalar as well as heavy vector pieces, the
leading order (in heavy meson mass $M$ and in number of derivatives of the
light fields) strong interaction is compactly written as [3]:
\begin{eqnarray}
\frac {1}{M} \cl_{\rm light~-~heavy} & = & iV_\m ~{\rm Tr}~ \{H[\partial_\m-
i\a\tilde{g}\rho_\m-i(1-\a)v_\m]\bar {H}\}\nonumber \\
               & + & id ~{\rm Tr}~ [H\g_\m\g_5p_\m\bar {H}]+\frac{ic}{m_{v}}
{}~{\rm Tr}~ [H\g_\m\g_\n F_{\m\n} (\r)\bar{H}],
\end{eqnarray}

\noindent wherein $m_{v}$ is the light vector meson mass introduced just to
keep
the coupling constant $c$ dimensionless and
\be
v_\m,p_\m= \frac{i}{2}(\xi \partial_\m\xi^\dagger \pm
\xi^\dagger\partial_\m\xi),
\ee

\noindent with the chiral matrix $\xi=\exp (i \phi/F_\pi)$.  Furthermore our
normalization convention sets $F_\p \simeq 132$ MeV.  $V_\m$ is the heavy meson
4-velocity and the heavy field $H$ here is taken to have the canonical
dimension of one. $\tilde{g}\simeq 3.93$ is the light vector-light
pseudoscalars coupling constant.  Heavy quark symmetry breaking terms, SU(3)
symmetry breaking terms as well as the chiral Lagrangian of the light sector
have not been explicitly written.

Notice that the light-heavy interaction (2.1) is characterized by the three
dimensionless coupling constants $\a$, $c$ and $d$ (denoted $g$ in [8]).
The choice $\a=1$ corresponds to a natural notion of light vector meson
dominance.  This choice sets to zero the coefficient of $ v_\m=
i/(2F^2_\p)(\phi
\partial_\m\phi-\partial_\m\phi\phi)+...~$  so that two pseudoscalars in a
p-wave state can only be emitted through an intermediate light vector particle
from a single heavy meson vertex.  Whether, in fact, $\a \approx 1$ remains to
be determined.

For our present application we also require the four fermion effective weak
interaction:
$$
\cl_W= \frac{G_{F}}{\sqrt{2}}J^{(+)}_\m J^{(-)}_\m \, ,
$$
$$
J^{(-)}_\m= i\bar{\n}_e\g_\m(1+\g_5)e +...\,,
$$
\be
J^{(+)}_\m = i V^*_{cs} \bar{s} \g_\m(1+\g_5)c + ... \, ,
\ee
\noindent with usual conventions [3] and where $V_{cs}$ is the
Kobayashi-Maskawa
matrix element.  The chiral covariant realization of the left handed hadronic
current, $J^{(+)}_{\m}$ in terms of heavy and light meson fields is
\be
J^{(+)}_\m/V^*_{cs}=F_D[\partial_\m D_b+i\a '\tilde g D_a\r_{\m ab}+
i(1-\a')D_a v_{\m ab} + MD^*_{\m b}](\xi^\dagger)_{b3} + ...\, ,
\ee
where the SU(3) triplet fields $(D_1,D_2,D_3)$ stand for $D^0,D^+D^+_s)$ and
similarly for the heavy vectors. (Eq. (2.4) is the same as  (4.6) of [3], but
we have redefined $(\a+\a ')$ by $\a '$ to avoid confusion). In (2.4) $\a '$ is
a
new dimensionless coupling constant (which scales however as $M$)
characterizing
the phenomenological hadron weak current.
We can rewrite (2.4) in the heavy quark limit as
\be
J^{(+)}_\m/V^*_{cs} = \frac {-iF_{D}M}{2} ~{\rm Tr}~
[\g_\m(1+\g_5)H_a](\xi^\dagger )_{a3}+ \frac{1}{2} F_D \a ' ~{\rm Tr}~
(\g_5H_a) (\tilde g \r_{\m ab} - v_{\m ab})(\xi^\dagger)_{b3}\, +... \,.
\ee

Now let us compute the hadronic matrix element for the process $D^0(p)\rt
K^-(p') +\pi^0(p'')+e^+(q_e)+\n_e(q_\n)$ We define the 4-momentum
\be
q=q_e+q_\n=p-p'-p''\, ,
\ee
\noindent and employ the following form factor decomposition:
$$\sqrt{8p_{0}p^\prime_{0}p^{\prime\prime}_{0}}\langle
K^-(p')\pi^0(p'')|J^{(+)}
_{\m}/V^{*}_{cs}|D^0(p)\rangle $$
\be
=-i[q_\m r+ (p'+p'')_\m\o_++(p'-p'')_\m\o_-
+h \e_{\m\a\r\n}p_\r p^{\prime}_\n p^{\prime\prime}_\a]\,.
\ee
\noindent The first term on the right hand side of (2.7) does not contribute to
the net weak amplitude since $q_\m$ dotted into the leptonic factor vanishes
for
zero lepton masses.  Hence the  form factor $r$ is irrelevant.  There are
contributions to the form factors both with and without intermediate light
vector meson $K^*$ poles.  The ``non-resonant'' (NR) diagrams without the
intermediate $K^*$ are discussed in the Appendix.  Taking the leading order in
$M$ contribution to each form factor yields
\begin{eqnarray}
(\o_+)_{NR} &=& \left (\frac {F_{D}}{2\sqrt{2}F^{2}_{\p}} \right )
\frac {dM}{\bigtriangleup - V\cdot p''}\, ,\nonumber \\
(\o_-)_{NR} &=& \left (\frac {F_{D}}{2\sqrt{2}F^{2}_{\p}} \right )
\left ( - \frac{dM}{\bigtriangleup-V\cdot p''}+ \a'  \right )\, ,\nonumber \\
h_{NR} &=& \frac{F_{D}d^{2}} {\sqrt{2 }F^{2}_{\p}}~~
\frac {1}{\bigtriangleup^{*}_{s}-V\cdot (p'+p'')}
{}~~\frac {1}{\bigtriangleup - V \cdot p''} \, ,
\end{eqnarray}

\noindent where $V_\m = p_\m/M$ is the 4-momentum of the  initial $D^0$,
$\bigtriangleup=M(D^*)-M(D)$ and $\bigtriangleup^{*}_{s}=M(D^{*}_{s})-M(D)$.
$\o_{+}$ and $\o_{-}$ both scale as $M^{1/2}$ while $h$ (since it multiplies
$p_\rho$) scales as $M^{-1/2}$.  Eq. (2.8) differs from an earlier calculation
[1] in the model with light {\it pseudoscalars}\, only by the $\a'$ term in
$(\o_-)_{NR}$.  $\a'$ is due to the presence of light vectors, as may be seen
from (2.4).  Chiral covariance demands an additional pseudoscalar piece when we
add the light vectors.  The analogous ``strong'' parameter $\a$ does not
appear in (2.8) but does show up in the form factor $r_{NR}$ given in (A7).

The computation of the diagrams containing $K^*$ poles can be simplified by
making use of earlier results on the $D\rt K^*$ weak current matrix element
[2-6].  From section 5 of [3] we obtain the leading large $M$ contribution in
this case as
\begin{eqnarray}
& &\sqrt{4p_{0}k_{0}} \langle K^{*-} (k,\e) | J^{(+)}_{\m}/V^{*}_{cs}
|D^0(p)\rangle\nonumber  \\
& =& \sqrt{4p_{0}k_{0}} \langle \bar{K}^{*0}(k,\e)|J^{+}_{\m}/V^*_{cs}|D^+(p)
\rangle \nonumber \\
&=& iF_D\bar\e_\n\left [ \a' \tilde {g}\d_{\m\n}+
\frac {2cM}{m_{v}}
\e_{\s\n\m\b}
{}~\frac {V_{\b}k_{\s}}{\bigtriangleup ^{*}_{s}-V\cdot k} -
\frac{\a \tilde{g}V_{\n}q_{\m}}{\bigtriangleup_{s}-V\cdot k} \right ] \, ,
\end{eqnarray}
\noindent wherein $q_\m=p_\m-k_\m$. Additional corrections due to higher
derivative interactions [3,4], loops [4] and excited heavy states [2,6] have
been discussed in the literature but (2.9) seems sufficient for our present
purpose.  The pieces of the $D^0(p)\rt K^-(p')+\pi^0(p'')$ transition matrix
element in (2.7) involving $K^*$ poles can be found from (2.9) simply by
defining
\be
k_\m=p^{\prime}_{\m}+p^{\prime \prime}_\m
\ee
\noindent and replacing $\bar {\e}_\n$ by the factor,
\be
\cf_\n= \frac{\tilde{g}}{2\sqrt{2}}\left ( \frac {m_{v}}{F_{\p}\tilde{g}}\right
)^2
\frac {\left [ p^{\prime}_{\n}-p^{\prime \prime}_{\n}-k_{\n}(m^{2}_{K}-m^{2}
_{\p})/m^{2}_{K^{*}}
\right ]}
{k^{2}+m^{2}_{K^{*}}-im_K^{*}\G_{K^{*}}} \,,
\ee
\noindent wherein the combination $(\frac {m_{v}}{F_{\pi}\tilde {g}})^2$ is
numerically
close to 2.0.  Eq. (2.11) is the product of the vector -2 pseudoscalar coupling
constant and the $K^*$ propagator.  Since the $K^*$ can go ``on shell'' we
include the conventional width correction in the denominator to maintain
unitarity.  Decomposing the ``resonant'' $K^*$ pole amplitudes (subscript R)
into the form factors defined in (2.7) gives:
\begin{eqnarray}
(\o_+)_R &=& - \frac {m^{2}_{K}-m^{2}_{\p}} {m^{2}_{K^{*}}}~  (\o_-)_R
\,, \nonumber \\
(\o_-)_R &=&-\left ( \frac {F_{D}} {2\sqrt {2} F^{2}_{\p}} \right )~
\frac {\a' m^{2}_{v}} {k^{2}+m^{2}_{K^{*}}-im_{K^{*}}\G_{K^{*}}}\, ,
\nonumber \\
h_{R} &=& \frac {2c\tilde {g}F_{D}} {\sqrt{2} m_{v}}~
\left ( \frac {m_{v}}
{F_{\p}\tilde{g}}\right )^2
{}~\frac {1} {\bigtriangleup^{*}_{s}-V\cdot (p'+p'')}~
\frac {1} {k^{2}+ m^{2}_{K^{*}}-im_{K^{*}}\G_{K^{*}}} \,
\end{eqnarray}

The net results for the $D^0\rt K^-\pi^0$ current transition form factors in
the limit of heavy charmed particles interacting with ``soft'' light
pseudoscalars and vectors are given by the sums of the corresponding terms
appearing in (2.8) and in (2.12).  The precise range of validity of the
concepts of a soft light pseudoscalar and (especially) a soft light vector are
probably best left to a comparison with experiment.  Certain results,  which
depend only on the existence of spontaneously broken chiral symmetry, are
designated ``current algebra theorems'' and should hold for zero mass
pseudoscalars as their 4-momenta go to zero.  In particular we may note that
$(\o_+)_R$ vanishes in this limit and that $(\o_-)_R$ (when we set
$m_{v}^{2}=m^{2}_{K^{*}}$ and $\G_{K^{*}}=0$, the latter corresponding to a
pure effective Lagrangian computation which thereby satisfies chiral symmetry
by construction) cancels off the $\a'$ piece of $(\o_-)_{NR}$.  In this
unphysical limit $\o_+=-\o_-$ [9].  We should stress that the fact that $\a'$
cancels out in this limit does not mean that it is not an important parameter
for describing the decay; actually, it turns out to be the most important
parameter.

To proceed with a comparison of the resonant and non-resonant contributions we
need at least a rough idea of the magnitudes of the ``strong'' parameters $\a,
c
$ and  $d$ as well as the ``weak'' parameter $\a'$.  Bounds on the value of $d$
have been obtained in the literature [10] which agree with a simple estimate
[3] based on pole dominance of the $D\rt K$ transition form factor:
\be
d\approx 0.53\,.
\ee
\noindent Information about $c$ and $\a'$ may be  obtained by comparing (2.9)
with
experimental information on the decays $D \rt K^*e^+\n_e$.  Eq. (2.9) is
expected to be most reliable for ``soft'' $K^*$'s, which implies that $(-q^2)$
should be as large as kinematically possible.  Now the experimental data is
analyzed in terms of form factors characterizing (2.9) which have the $q^2$
dependence, $M^{*2}_{s}/(M^{*2}_{s}+q^2)$.  Evaluating (2.9) at the extreme
value $-q^2=(M-m_{K^{*}})^2$, {\it assuming\/} the above $q^2$ dependence to
extrapolate to small $-q^2$ and comparing with the experimental values [11]
yields the estimates
\be
|\a'|\approx 1.73,\sp |c|\approx 1.6\, .
\ee
\noindent $\a$ is not determined from this analysis.  However a study of
the binding energy of
heavy baryons as solitons of the meson Lagrangian involving (2.1) shows that
(5.3) of Ref. [12] may be fit with the choice
\be
d=0.53,\sp c=1.6,\sp \a=-2\,.
\ee
\noindent We stress that the numerical estimates (2.13)-(2.15) are preliminary
in nature; however they should suffice to draw qualitative conclusions.

\sxn{Comparison of Light Pseudoscalar and
 Light Vector Amplitudes}

First we must discuss the kinematics associated with the hadronic matrix
element (2.7).  For the decay $D^0(p)\rt
K^-(p')+\pi^0(p'')+e^+(q_e)+\n_e(q_\n)$
there are five independent dynamical variables needed to specify the momentum
configuration; a recent discussion is given in Ref. [1] (see also [13]) based
on earlier treatments [14] of $k_{e4}$ decays.  It is somewhat simpler if we
confine our attention just to the hadronic part (2.7).  Then, for each value of
the invariant lepton squared mass $-q^2$ we have in effect  a three body final
state.  The classic Dalitz plot analysis shows that the kaon energy $E'$ and
the pion energy $E''$ (both in the $D^0$ rest frame) are sufficient.
Altogether this gives $E',E''$ and $q^2$ as a possible complete set of
variables to describe (2.7).  Since $k^2$ plays an important role in the $K^*$
pole amplitude we shall choose the alternative set
\be
(k^2, q^2, E'')\,.
\ee
\noindent A convenient formula relating the two sets is
\be
E'+E''= \frac {M^2-k^2+q^2}
{2M} \equiv Q (k^2, q^2)\, ,
\ee
\noindent where $M$ is the $D^0$ mass.  With the neglect of the lepton masses,
$-k^2$ satisfies
\be
(m_\p+m_k)^2 \leq -k^2 \leq M^2 \, .
\ee
\noindent For a given $k^2$, $q^2$ must satisfy
\be
0 \leq - q^2 \leq (M-\sqrt{-k^{2}})^2 \, .
\ee
\noindent  The region defined by (3.3) and (3.4) is illustrated in Fig. 1.
Now, for each value of $q^2$ there is a Dalitz plot boundary in the
$k^2-E''$ plane obtained by rewriting the condition $|\bf {p'}\cdot
\bf {p''}|=|\bf {p'}||\bf {p''}|$ in terms of the set (3.1).  The
allowed regions, corresponding to horizontal slices perpendicular to the page
in Fig. 1, are illustrated in Fig. 2. A simple formula can be given
for the ``average'' value of $E''$ at each $(k^2,q^2)$:
\begin{eqnarray}
\bar E''(k^2,q^2) &=& \frac {1}{2} \left [ E''_{\rm max} (k^2, q^2)+ E''_{\rm
min}(k^2, q^2) \right ] \nonumber \\
&=& \frac {1}{2} Q (k^2,q^2) \left [ 1 +
\frac {m^{2}_{K}-m^{2}_{\p}} {k^{2}} \right ].
\end{eqnarray}
\noindent Notice that Fig. 1 does not correspond to a {\it fixed\/} $E''$ slice
but rather represents the {\it projection\/} of the $(k^2,q^2,E'')$ boundary
surface into
the $(k^2,q^2)$ plane.

The expressions for the form factors in the heavy quark limit given by (2.8) +
(2.12) are expected to be most reliable for large $(-q^2)$, corresponding to
``soft'' $\pi^0$ and $K^-$ particles as well as, in the spirit of the present
Lagrangian, soft $K^*$ particles.  Refering to Fig. 1, we see that the soft
$K^*$ condition suggests that we consider the effective Lagrangian expressions
to hold in the region where $-q^2$ is greater than around 0.5 ${\rm GeV}^2$.
This represents a fairly large portion of the entire phase space.  Nevertheless
it is not unreasonable from an experimental point of view inasmuch as the
Particle Data Group tables state [15] that ``it is generally agreed that the
$\bar {K} \pi e^+ \n_e$ decays of the $D^\dagger$ and $D^0$ are dominantly
$\bar{K}^*e^+\n_e$.'' Experimentally, it is also
known that the amplitudes are damped for decreasing $(-q^2)$; this can be seen
from Fig. 1 to decrease the importance of the larger $-k^2$ region.

{}From a theoretical point of view it is very interesting to compare the
pseudoscalar and vector contributions to the hadronic form factors. Nature
tells us, of course, that the vector contributions dominate when one folds in
the leptons and integrates their ``squares'' over all phase space.
Nevertheless, it is important to get an idea of the ``local'' ratios of the two
contributions in various kinematic regions.  The phenomenological analysis of
Ref. [16], for example, shows that it is the form factor proportional to
$\bar{\e}_\m$ in (2.9) (i.e, the $\a'$ term) which mainly supplies the total
width for $D^0\rt K^{*-}e^+\bar {\nu}_e$.  This form factor contributes to both
$(\o_+)_R$ and $(\o_-)_R$ in (2.12) but $(\o_-)_R$ is clearly the dominat one;
$(\o_+)_R$ would vanish when the
$k$ and the $\p$ masses are equal or neglected.  Hence we will focus our
attention on the $\o_-$ form factor.  It is convenient to  rewrite it as
follows (in
the $D^0$ rest frame):
\be
\o_- = \frac {F_{D}} {2 \sqrt {2} F^{2}_{\pi}}
\left [ \a' \left (1- ~\frac
{m^{2}_{K^{*}}} {k^{2}+m^{2}_{K^{*}} - im_{K^{*}}\G_{K^{*}}}\right)
-~\frac {dM} {\bigtriangleup + E''} \right ].
\ee
\noindent We noted earlier that the coefficient of the $\a'$ term (which
corresponds to the extra piece introduced by adding light vectors to the
effective Lagrangian) vanishes in the unphysical $k_\m \rt 0$ limit (with
$\G_{K^{*}}=0$) in agreement with expectations.  Let us then consider the ratio
of the magnitude of the $\a'$ term to the magnitude of the last term:
\be
{\rm {ratio}} = \left |\frac {\a'}{d}\right | ~\frac {(\bigtriangleup + E'')}
{M}
\left [ \frac {k^{4}+m^{2}_{K^{*}} \G^{2}_{K^{*}}}
{(k^2 +m^{2}_{K^{*}})^{2} +m^{2}_{K^{*}}\G^{2}_{K^{*}}} \right ]^{1/2}.
\ee
\noindent Only $\a'$ and $d$ are not very well known; we will use the estimates
in
(2.13) and (2.14) which are certainly qualitatively reasonable.  There is a
lower bound for the above,
\be
{\rm {ratio}} \geq \left |\frac {\a'}{d} \right |~\frac {\bigtriangleup
+m_{\p}}{M}
{}~\frac {(m_{K} + m_{\pi})^{2}}{m^{2}_{K^{*}}- (m_{K}+m_{\pi})^{2}} \approx
0.5,
\ee
\noindent where $\G_{K^{*}}$ was set to zero for simplicity since it has a
negligible effect when $(-k^2)$ is as small as possible.  This result indicates
that there is no region in which the vector contribution is negligible compared
to the pseudoscalar contribution.  So if one were to make the usual derivative
expansion of the chiral perturbation theory approach, there would be large
corrections to the first order results due to the existence of the $K^*$. Of
course, the pseudoscalar piece takes on its largest value near the cusp in Fig.
1.

Since (3.7) depends only on $k^2$ and $E''$ it is convenient to display curves
of
constant ratio in Fig. 2.  We notice that there is only a very small region for
which the ratio is less than unity.  Because of the small width of the $K^*$
(50
MeV) the ratio rises dramatically to around 15 at $-k^2=m^{2}_{K^{*}}$.  It is
seen that the large ratio of amplitudes persists over a non-negligible region.

The ratio $\left |\frac {h_{R}}{h_{NR}}\right |$ of the magnitudes of the
vector and
pseudoscalar contributions to the weak vector current form factor is seen from
(2.8)  and (2.12) to be the same as the ratio (3.7) when we multiply the latter
by the factor
\be
2 \left | \frac {c} {\a'\tilde {g}}\right | ~
\frac {m_{K^{*}}M} {(k^{4}+m^{2}_{K^{*}}\G^{2}_{K^{*}})^{1/2}}\, .
\ee
\noindent This is numerically around 1.0 in the phase space region of interest
so the $K^{*}$ contribution is also dominant for
this form factor.

Tu sum up, in the large $(-q^{2})$ region (optimistically as large as
$-q^2\geq
0.5 ~ GeV^2$) the $K^*$ contribution overwhelms the pseudoscalar
contribution nearly everywhere.  Even for the very largest $(-q^2)$, where the
soft pseudoscalar results are expected to be most significant, the $K^*$
amplitudes are relatively sizeable.  This situation would seem to suggest the
desirability of a modified chiral perturbation theory program in which both
pseudoscalars and vectors are retained in the effective Lagrangian from the
very beginning.  Some recent discussion of this point of view has been given in
Ref. [17].  In the case where one is considering a non-strange transition
matrix element (like $B \rt \p\p)$ we should replace in (3.8), $(m_K+m_\p)$ by
$2 m_\p$  and $m_{K^{*}}$by $m_\r$.  Then there would be a very small region in
which the pseudoscalar piece might be considered dominant, but the overall
picture would be qualitatively similar.

\sxn{Remarks}
1.~~Of course, it would be a better approximation to deal with the $B$ rather
than the $D$ meson as an example of a heavy field.  Similarly it would be a
better approximation to restrict the light particles to the non-strange ones.
Thus a similar analysis (with basically identical formulas) would be somewhat
cleaner for $\bar {B} \rt \pi\pi$ current  transitions rather than $D\rt
\bar{K}\pi$ transitions.  Nevertheless there is at present much more
experimental data for the $D\rt \bar{K}\p$ case and  it is quite identeresting
in its own right.

2.~~As noted after (2.9), the formula we are using for the resonant
contribution neglects a number of effects which are subleading from the light
meson point of  view but may be necessary to take into account if the
relatively small form factor which would be proportional to $p_\m$ in (2.9) is
actually non-zero.  However we expect  that our approximation is sufficient for
the purpose of comparing the relative strengths of the vector and pseudoscalar
contributions.

3.~~The ratio of the non-resonant part of $\G(D\rt\bar{K}\p e^+\n_e)$ to
$\G(D\rt \bar{K}^*e^+\n_e)$ is of evident experimental interest. The precise
meaning of this quantity depends on the manner in which it is extracted from
experiment.  Apparently, there is  no universal method.  The most
straightforward way is simply to define the resonant contribution as everything
within a certain band of $-k^2$ surrounding $m^2_{K^{*}}$.  This definition
has, however, the misleading feature that is counts non-resonant background
near the peak as resonant.  It is particularly easy to apply this definition to
the present case when one makes the reasonable approximation that the {\it
entire} amplitude is dominated by the $K^*$ pole diagrams.  As recently
illustrated for a different decay in Ref. [13] both the phase space and the
squared amplitudes factorize in this approximation so that one obtains
\be
\G(D^0\rt K^-\p^0 e^+\n_e)\approx \frac
{\G(K^{*-}\rt K^{-}\p^{0})}
{\G_{K^{*}}}
{}~\G(D^0\rt K^{*-}e^+\n_e)\frac {1}{\p} ~
\int_{-\infty}^{\infty}~\frac {dx}{x^{2}+1},
\ee
\noindent where $x$ is defined by $k^2+m^{2}_{K^{*}}=x \G_{K^{*}}m_{K^{*}}$.
If
the ``resonant'' region is taken to be that range of $-k^2$ for which
$|\sqrt{-k^{2}}-m_{K^{*}}|<N\G_{K^{*}}$ then integrating (4.1) yields the
non-resonant/resonant ratio to be about
\be
\frac {\p}{2~{\rm tan}^{-1}(2N)} -1 .
\ee
This gives a 19\% non resonant contribution for $N=2$ and 12\% for $N=3$.  The
extent to which the data obeys (4.2) as a function of $N$ might be considered a
measure of how good is the $K^*$ dominance.  The fact that we do get a
non-resonant contribution at all with pole dominance is, of course, an artifact
of its definition.  However, it also illustrates the difficulty in giving a
meaningful {\it experimental} definition of the non-resonant/resonant ratio.
Since the ``theoretical'' resonant and non-resonant amplitudes have the same
order of magnitude {\it outside} the resonance region, the above numbers may
give a satisfactory rough order of magnitude estimate for any reasonable
definition of the ratio.
\vglue 1.0cm
\centerline{\bf Acknowledgements}
\vglue 0.5cm

This work was supported in part by the U.S. Department of Energy under contract
number DE-FG-02-85ER40231.
\newpage
\centerline {\bf Appendix}
\vglue 0.5cm

For the process $D^0\rt K^-\p^0e^+\n_e$ we require the hadronic matrix element
$$
\sqrt{8 p_0p^{\prime}_{0}p^{\prime\prime}_{0}} \langle
K^-(p')\p^0(p'')|J^{+}_{\m}/V^{*}_{cs}|D_0(p)\rangle, \eqno(A1)
$$
where the states are normalized in a unit volume.  We shall proceed by first
using the ordinary, rather than ``heavy'', meson fields to compute (A1)
(see section V of Ref.[3]) and then
take the heavy quark limit.  The contribution to (A1) from the contact
(non-pole) diagram is:
$$
\frac {-iF_{D}} {2\sqrt{2}F^{2}_{\p}}
[p_\m+(\a'-1)(p^{\prime}_{\m}-p^{\prime\prime}_{\m})]. \eqno (A2)
$$
\noindent The $D^{+}_{s}$ pole diagram, wherein $D^0(p)\rt
K^-(p')+\p^0(p'')+D^{+}_{s}(q)$ followed by $D^{+}_{s}(q)\rt e^+\n_e$,
contributes the term
$$
\frac {iF_{D}(1-\a)} {2\sqrt{2}F^{2}_{\p}} ~
\frac {(p^{'}-p^{''})\cdot (p+q)} {M^{2}_{s}+q^{2}} q_\m\, , \eqno (A3)
$$
in which $M_s$ is the $D^{+}_{s}$ mass and $q_\m\equiv p_\m-(p'+p'')_\m$.
The $D^{*0}$ pole diagram, wherein $D^0 \rt \p^0 +D^{*0}$ at the strong vertex
followed by $D^{*0}\rt K^-+\m^+\n_\m$ at the weak vertex, contributes:
$$
\frac {-i \sqrt{2}M^{2}F_{D}d}{F^{2}_{\p}}~~
\frac {[p^{\prime\prime}_{\m}+(p-p'')_{\m}(p-p^{\prime\prime}) \cdot
p^{\prime\prime}/M^{*2}]}
{(p-p^{\prime\prime})^{2} + M^{*2}}\,  , \eqno(A4)
$$
\noindent in which $M^{*}$ is the $D^{*0}$ mass.  There are also two
double-pole diagrams.  The first features $D^0\rt \p^0+D^{*0}$ at a strong
vertex followed by $D^{*0} \rt K^-+D^{*+}_{s}$ at another strong vertex which,
in turn, is followed by $D^{*+}_{s}\rt e^+\n_e$ at the weak vertex.  The
contribution to (A1) is
$$
\frac {-i2\sqrt{2}d^{2}M^{2}F_{D}} {F^{2}_{\p}} ~
\frac {\e_{\m\a\r\n} p_{\r}p^{\prime}_{\n}p^{\prime\prime}_{\a}}
{[M^{*2}_{s} + q^{2}][M^{*2}+(p-p^{\prime\prime})^{2}]}, \eqno(A5)
$$
\noindent in which $M^{*}_{s}$ is the $D^{*+}_{s}$ mass. Finally, the
diagram with $D^0\rt \p^0+D^{*0}, D^{*0}\rt K^-+D^{+}_{s},
D^{+}_{s}\rt e^{+}\n_e$ gives
$$
\frac {-i2\sqrt{2}F_{D}M^{2}d^{2}} {F^{2}_{\p}} ~
\frac {q_{\m}[p'\cdot p'' + p' \cdot (p-p'')p''\cdot(p-p'')/M^{*2}]}
{[q^{2}+M^{2}_{s}][(p-p'')^{2}+M^{*2}]}\, . \eqno(A6)
$$
\noindent In order to obtain a model independent result corresponding to
$M\rt\infty$ we should delete terms which fall off faster than $M^{1/2}$, the
scale behavior of (A1).  For example, using $F_D\sim M^{-1/2}$, and $\a'\sim
M$, (A2) $\rt$
$$
\frac{-iF_{D}}
{2\sqrt{2}F^{2}_{\p}}
[p_\m+\a' (p'-p'')_\m ]. \eqno (A2')
$$
\noindent In taking the limit of (A3) we set $p_\m=MV_{\m}$ and throw away
terms of quadratic order in $p'$ and $p''$.  The resonance denominator of (A3)
becomes $2M [\bigtriangleup_{s} - V \cdot(p'+p'')]$ with
$\bigtriangleup_{s}=M_s-M$ and (A3) $\rt$
$$
\frac {-i F_{D}(1-\a)}
{2\sqrt{2}F^{2}_{\p}}
{}~\frac {V\cdot(p'-p'')} {\bigtriangleup_{s}-V\cdot(p'+p'')} q_\m \, .
\eqno (A3')
$$
\noindent The other diagrams are treated similarly.  With the form factor
decomposition defined in (2.7) we obtain the results listed in (2.8) for the
three experimentally significant form factors.  For the sake of completeness we
also give here:
$$
r_{NR}= \frac {F_{D}} {\sqrt{2}F^{2}_{\p}}
\left [ \frac {1}{2} +
\frac {(1-\a)}{2F^{2}_{\p}}
{}~\frac {V\cdot(p'-p'')} {\bigtriangleup_{s}-V\cdot (p'+p'')}
+ \frac{dV\cdot p''}{\bigtriangleup-V\cdot p''}
{}~+\frac {d^{2}(p'\cdot p''+V\cdot p'V\cdot p'')}
 {[\bigtriangleup_{s}-V\cdot(p'+p'')](\bigtriangleup-V\cdot p'')} \right ].
\eqno(A7)
$$

\newpage

\centerline {\bf References}
\vglue 0.5cm
\begin{enumerate}
\item C. Lee, M. Lu and M. Wise, {\it Phys. Rev.} {\bf D46}, 5040 (1992).

\item R. Casalbuoni, A. Deandrea, N. Di Bartolomeo, R. Gatto, F. Feruglio and
G. Nardulli, {\it Phys. Lett.} {\bf 292}, 371 (1992); {\bf 299}, 139 (1993).

\item J. Schechter and A. Subbaraman, {\it Phys. Rev.} {\bf D48}, 332 (1993).

\item P. Ko, {\it Phys. Rev.} {\bf D47}, 1964 (1993).

\item N. Kitazawa and T. Kurimoto, Osaka Report no. OS-GE-27-92
(hep-ph/9302296).

\item A. Kamal and Q. Xu, Alberta report - Thy-31-93.

\item A recent review is given in Ulf-G. Meissner Bern University report
BUTP-93/01.

\item M. Wise, {\it Phys. Rev.} {\bf D45}, R2188 (1992); T. Yan, H. Cheng, C.
Cheung, G. Lin, Y. Lin and H. Yu, {\it Phys. Rev} {\bf D46}, 1148 (1992); G.
Burdman and J. Donoghue, {\it Phys. Lett.} {\bf B280}, 280 (1992); P. Cho {\it
ibid} {\bf 285}, 145 (1992); A. Falk and M. Luke {\it ibid} {\bf 292}, 119.

\item For a review, see section 5.4 D of R. Marshak, Riazuddin and C.P. Ryan,
{\it Theory of Weak Interactions in Particle Physics}\/ Interscience, 1969.

\item In addition to [8] above see P. Cho and H. Georgi, {\it Phys. Lett.} {\bf
B296}, 408 (1992); J. Amundsen et al. {\it ibid} {\bf 296}, 415 (1992).

\item K. Kodama et al., {\it Phys. Rev. Lett.} {\bf 66}, 1819 (1991); {\it
Phys. Lett.} {\bf B263}, 573 (1991); {\it ibid} {\bf 286}, 187 (1992);
J. Anjos et al., {\it Phys. Rev. Lett.} {\bf 62}, 1587 (1989);
{\it ibid} {\bf 65}, 2630 (1990); {\it ibid} {\bf 67}, 1507 (1991).

\item K.S. Gupta, M.A. Momen, J. Schechter and A. Subbaraman, {\it Phys. Rev.}
{\bf D47}, R4835 (1993).

\item H-Y Cheng et al., Cornell University report CLNS 93/1204.

\item A. Pais and S. Trieman, {\it Phys. Rev.} {\bf 168}, 1858 (1968); N.
Cabibbo and A. Maksymowicz {\it ibid} {\bf 137}, 13438 (1965).

\item See note $v$ following the meson summary table in Particle Data Group, K.
Hikasa et al., {\it Phys. Rev.} {\bf D45}, 51 (1992).

\item X.-Y. Pham {\it Phys. Rev.} {\bf D46}, 1909 (1992); {\bf 47} 350 (E)
(1993).

\item J. Schechter, A. Subbaraman, and H. Weigel, {\it Phys. Rev} {\bf 48} 339
(1993).
\end{enumerate}

\vglue 0.6cm
\centerline {\bf Figure Captions}
\vglue 0.5cm

\begin{itemize}
\item [\bf {Fig. 1}]  Projection of the three dimensional $(k^2, q^2, E'')$
phase
space boundary into the $(k^2,q^2)$ plane.

\item [\bf {Fig. 2}] Phase space boundaries in the $k^2-E''$ plane at various
values
of $q^2$. In increasing order of size the closed curves correspond respectively
to $-q^2$ = 0.925, 0.725, 0.525, 0.325 and 0.025 GeV$^2$.  Also shown are
points on the contour lines on which the ratio in (3.7) takes on fixed values;
the
circles, crosses and squares correspond respectively to the ratio equal to 1,3
and 9.

\end{itemize}
\end{document}